\documentclass[11pt]{article}
\usepackage[colorlinks=true,citecolor=blue,linkcolor=blue]{hyperref}
\usepackage{multirow}
\usepackage[left=2cm,right=2cm,top=2cm,bottom=2cm]{geometry}
\usepackage{cite}
\usepackage{psfrag}
\usepackage{graphicx}
\usepackage{graphics}
\usepackage{epsfig}
\usepackage{amsmath,amsfonts,amssymb}
\usepackage{pstcol,pst-fill,pst-grad}
\usepackage{euscript}
\usepackage{pstricks}
\usepackage{wrapfig}
\usepackage{slashed}
\usepackage[toc,page]{appendix}
\usepackage{here}

\date{}
\begin{document}
	\title{\vspace{-3cm}
		\hfill\parbox{4cm}{\normalsize \emph{}}\\
		\vspace{1cm}
		{Two-body hadronic decay of $K^{+}$ in the presence of a circularly polarized laser field}}
	\vspace{2cm}
	
	\author{M Baouahi$^{1}$, I Dahiri$^{1}$, M Ouali$^{1}$, B Manaut$^{1}$, R Benbrik$^{2}$ and S Taj$^{1}$\thanks{Corresponding author, E-mail: s.taj@usms.ma}  \\
		{\it {\small$^1$ Recherche Laboratory in Physics and Engineering Sciences, Team of Modern}}\\
		{\it {\small and Applied Physics, FPBM, USMS, Morocco}}\\
		{\it {\small$^2$ LPFAS, Polydisciplinary Faculty of Safi, UCAM, Morocco.}}\\			
	}
	\maketitle \setcounter{page}{1}
\date{\today}
\begin{abstract}
In this study, we have investigated the two-body hadronic decay of the charged kaon, $ K^{+}\rightarrow\pi^{+}+\pi^{0} $, in the presence of a laser field with circular polarization. We have derived, by analytical techniques, the laser-assisted decay width and the branching ratio of the charged kaon decay via the two-body hadronic channel. We have also taken into consideration the impressive results obtained for the laser-assisted charged kaon decay via the leptonic mode in order to understand more clearly the effect of the laser field on the quantities related to the charged kaon decay such as the decay width, the branching ratio and lifetime.
A precise comparison of the ratios of hadronic to muonic decay in the presence of the laser field is made to show that the hadronic mode becomes slightly more important by increasing the laser field intensity.
\end{abstract}
\section{Introduction}
The interaction between the electromagnetic field and charged particles is a very interesting research topic in the study of particle physics processes, especially after the introduction of the laser which is one of the most important technological discovery in 1960 by Maiman\cite{Maiman:1960}.
Today, the breakthrough progress in laser technology allow us to observe in real time the electrons and atoms that compose matter, and to deform, in an ultra-fast way, the molecular structure around the atoms in order to transform materials into magnetic materials\cite{Cammarata:2021}. Moreover some studies and discoveries, have been made in 2018, have demonstrated that the energy of sub-relativistic electrons is strongly modified on the scale of a few femtoseconds by the interaction with a progressive wave created in the vacuum by the collision of two laser pulses of different frequencies\cite{Kozak:2018}.

Studies in the field of laser-matter interactions usually deal with non-relativistic \cite{Joachain:1994} and relativistic atomic physics \cite{Manaut1:2004,Taj:2005}. For example, in the study of Mott scattering of an electron by a charged nucleus in the presence of an electromagnetic field with a given polarization (linear, circular or elliptical), they showed that the intensity of the electric field, the kinetics of the incident electron and the collision geometry influence the behavior of the cross section\cite{Manaut:2004}. Today, the technological progress gives us the opportunity to introduce an external electromagnetic field in high energy physics interactions. Indeed, several studies have been made to study the impact of the laser field on the decay width\cite{Nikishov:1964,Akhmedov:1983,Reiss:1983,Becker:1984,Baouahi:2021} and on the effective cross section \cite{Manaut:2004,Muller:2004,Taj:2012,all:2021} by introducing the phenomenon of emission and/or absorption of photons\cite{Volkov:1935}. For instance, in the study of nuclear decay $\beta$, there have been debates on the possibility of having significant effects by the presence of a strong laser pulse \cite{Nikishov:1964,Akhmedov:1983,Reiss:1983,Becker:1984}. In \cite{Mouslih:2020,Jakha:2021}, the authors illustrated the effect of an electromagnetic field with circular polarization on the decay widths and lifetime of the charged pion and the $Z$ boson. In \cite{Ouali:2021}, it is found that the circularly polarized laser field has a great impact on the cross section of the charged Higgs pair production via electron-positron annihilation. These processes are called laser-assisted processes for that they may occur also in the presence of the electromagnetic field.

There is another category of very well known processes in laser physics, and they are induced by the laser field. We cite, here, some examples in which the creation of pairs $\{ e^{+},e^{-}\}$ is induced by the presence of an external field. In \cite{Sauter:1931,Schwinger:1951,Brezin:1970}, the authors showed that the creation of electron-positron pairs is possible by a strong plane electromagnetic wave in the presence of an additional electromagnetic energy source. In \cite{Breit:1934}, the pairs of $\{ e^{+},e^{-}\}$ are produced when a high energy photon of wave four-vector $ k'^{\mu} $ is propagating in a strong laser field of wave four-vector $ k^{\mu}$. However, the pair production, in this case, is possible only under the condition that $n(k'k)>2m_{e}$ with $n$ representing the minimal integer for this condition to be valid. Another example of electron-positron pair production is illustrated in \cite{Bell:2008,Kirk:2009} where two strong laser beams propagate in opposite directions and generate a stationary light wave. For all these studies, the production of ultrashort pulses is important, not only because time compression obviously implies an increase of the intensity at a given laser energy, but also because high intensities allow, in general, to control fast physical processes\cite{Mourou:2011,Piazza:2012}.

In our previous work \cite{Baouahi:2021}, we have found impressive results on the laser-assisted charged kaon decay in the leptonic channel. In addition, we have indicated that the $CPT$ symmetry might be preserved or broken in both the muonic and the electronic channels. This is due to the fact that the parameter associated with the $CPT$ symmetry is affected by the circularly polarized laser field. Therefore, we can control the dominance of matter over antimatter or vice versa. In this respect, we have investigated the two-body hadronic decay of the charged kaon in the presence of an electromagnetic field of circular polarization by using the same laser parameters as in \cite{Baouahi:2021}. Thus, the aim of this work is to present the analytical and numerical results about the laser-assisted decay of the charged kaon, which has free a lifetime $ \tau=(1.2380\pm0.0020)\times10^{-8}\,s $\cite{PDG:2020}. Indeed, we have illustrated the effect of the circularly polarized laser field on different quantities associated to the two-body hadronic decay of the positive kaon.

The remainder of this work is organized as follows: In the next section (\ref{Sec.1}), we will deal with the theoretical and analytical calculation of the charged kaon decay width in the absence and presence of a circularly polarized laser field. Then, we will present in section (\ref{Sec2}) the obtained data and results. A short conclusion is given in section (\ref{Sec3}). Some theoretical details of the positive kaon decay width calculations are given in the appendix. We mention that, in the laboratory system, this disintegration study is represented in the system of natural units $ \hbar=c=1 $, and the signature metric of space-time is chosen as $ (+---) $.
\section{Theoretical framework}\label{Sec.1}
In this part, we will start with the theoretical calculation of the laser-free two-body hadronic decay of the charged kaon, $ K^{+}$, at the lowest order without taking into account the quarks composition of mesons. In order to study and analyze the effect of an electromagnetic field on this decay, the second section will deal with the decay process by dressing the charged mesons.
The plane wave functions used to describe the neutral meson, $\pi^{0}$, and charged mesons that are involved in the studied decay, in the absence of the electromagnetic field, are derived from the free Klein-Gordon equation. However, in the presence of an electromagnetic field, the charged particle are described by the Volkov functions\cite{Volkov:1935}.
\subsection{Laser-free kaon decay}
 In the absence of an external field, the transition matrix element $S_{fi}$ which is associated with the decay of the charged kaon $ K^{+}$ to $\{\pi^{+},\pi^{0}\}$ is written as a product of two currents such that:
\begin{eqnarray}\label{Sfi_Transition_1}
S_{fi}=\dfrac{-iG_{F}}{\sqrt{2}}\int d^{4}x\mathcal{J}_{\mu}^{\langle\,K^{+}\vert\,\mathcal{O}_{1}\vert\pi^{0}\rangle\dagger}(x)\times\mathcal{J}^{{\langle\,0\vert\,\mathcal{O}_{2}\vert\,\pi^{+}\rangle}\mu}(x),
\end{eqnarray}
with $ G_{F} $ represents the Fermi constant. The two hadronic currents $ \mathcal{J} $  \cite{Hayashi:1966,Okun:1984} are defined as follows:
\begin{eqnarray}\label{Currents_1}
\mathcal{J}_{\mu}^{\langle\,K^{+}\vert\,\mathcal{O}_{1}\vert\pi^{0}\rangle}(x)&=&\dfrac{1}{2V\sqrt{E_{1}E_{3}}}\langle\,K^{+}\vert\,\mathcal{O}_{1}\vert\pi^{0}\rangle\,e^{i(P_{1}-P_{3}).x}\\
&=&\dfrac{f_{+}(\Delta)[P_{1}+P_{3}]_{\mu}+f_{-}(\Delta)[P_{1}-P_{3}]_{\mu}}{2V\sqrt{E_{1}E_{3}}}\,e^{i(P_{1}-P_{3}).x},\\
\mathcal{J}^{{\langle\,0\vert\,\mathcal{O}_{2}\vert\,\pi^{+}\rangle}\mu}(x)&=&\dfrac{1}{\sqrt{2E_{2}V}}\langle\,0\vert\,\mathcal{O}_{2}\vert\pi^{+}\rangle\,e^{iP_{2}.x}\\
&=&\dfrac{iP_{2}^{\mu}F_{\pi^{+}}}{\sqrt{2E_{2}V}}\,e^{iP_{2}.x},
\end{eqnarray}
where $ V $, $ P_{i}=(E_{i},\vec{p_{i}}) $ ($ i=\{1,2,3\} $) and $ \Delta $ are respectively the quantum volume, the 4-momentum of the hadrons (successively for $ K^{+},\, \pi^{+} $ and $ \pi^{0} $) and the transfer momentum such that $ \Delta=(P_{1}-P_{3})^{2} $. $F_{\pi^{+}} $ is the decay constant associated with the particle $ \pi^{+} $. The operator $ \mathcal{O}_{1} $ corresponds to the transition in the vector current $ [K^{+}, \pi^{0}] $, and $ \mathcal{O}_{2} $ corresponds to that in the weak vector-axial current. The form factors $ f_{\pm} $ are given, as in the case of $ K^{+}_{e3} $ and $ K^{+}_{\mu3} $ decays, by the following expression\cite{PDG:2020}:
\begin{eqnarray}\label{Forme_factors_1}
f_{\pm}(\Delta)=f_{\pm}(0)\left[ 1+\lambda_{\pm}\dfrac{\Delta}{m_{\pi^{+}}^{2}} \right].
\end{eqnarray}
We use the parameterization $ \{\lambda_{+},\lambda_{0}\} $ to define the factors $ f_{\pm} $ by the function $ f_{0} $ \cite{PDG:2020} which is given by:
\begin{eqnarray}\label{Forme_factors_2}
f_{0}(\Delta)=f_{+}(\Delta)+\dfrac{\Delta}{m_{K^{+}}^{2}-m_{\pi^{0}}^{2}} f_{-}(\Delta) \qquad \text{with} \qquad f_{0}(\Delta)=f_{0}(0)\left[ 1+\lambda_{0}\dfrac{\Delta}{m_{\pi^{+}}^{2}} \right].
\end{eqnarray}
In this parameterization, we use the universality assumption $\mu-e$ to define $ \lambda_{+} $ and $ \lambda_{0} $\cite{PDG:2020}.
After weighting the square of the matrix element $ S_{fi} $ by the phase space and per unit time $ T $, the decay width \cite{Greiner:2000} becomes as follows:
\begin{eqnarray}\label{Decay_1}
\Gamma &=& \dfrac{1}{T}\int \dfrac{Vd^{3}\vec{P}_{2}}{(2\pi)^{3}}\int \dfrac{Vd^{3}\vec{P}_{3}}{(2\pi)^{3}}\vert \bar{S}_{fi}\vert^{2}\nonumber\\
&=& \dfrac{G_{F}^{2}F_{\pi^{+}}^{2}}{32\pi m_{K^{+}}}\sqrt{\dfrac{m_{K^{+}}^{4}+m_{\pi^{+}}^{4}+m_{\pi^{0}}^{4}-2m_{K^{+}}^{2}m_{\pi^{+}}^{2}-2m_{K^{+}}^{2}m_{\pi^{0}}^{2}-2m_{\pi^{+}}^{2}m_{\pi^{0}}^{2}}{(m_{K^{+}}^{2}-m_{\pi^{+}}^{2}+m_{\pi^{0}}^{2})^{2}}} \nonumber\\
&&\left[ (m_{K^{+}}^{2}-m_{\pi^{0}}^{2})f_{+}(m_{\pi^{+}}^{2})+m_{\pi^{+}}^{2}f_{-}(m_{\pi^{+}}^{2}) \right]^{2}.
\end{eqnarray}
\subsection{Laser-assisted kaon decay}
An electromagnetic field with a circular polarization is described by the classical 4-vector potential $ A^{\mu}(\phi)=a_{1}^{\mu}\cos(\phi)+a_{2}^{\mu}\sin(\phi) $, where $ \phi=k.x $ is its phase. The 4-vector $ A^{\mu} $ verifies the transversality condition $ k.A=0 $ (Lorentz gauge), with $ k=(\omega,\vec{k}) $ is the 4-vector wave. The quadri-polarizations $ a_{1}^{\mu}=|a|(0,1,0,0) $ and $ a_{2}^{\mu}=\vert a\vert|(0,0,1, 0) $ are equal in magnitude such that $ a_{1}^{2}=a_{2}^{2}=a^{2}=-\vert a\vert^{2}=-(\varepsilon_{0}/\omega)^{2} $, and they verify the orthogonality relation $ a_{1} a_{2}=0 $. The quantity $ \varepsilon_{0} $ represents the amplitude of the electric field of the laser while $ \omega $ is its frequency.
By adding the formalism of the laser field to describe the decay $ K^{+}\rightarrow\pi^{+}+\pi^{0} $, the Klein-Gordon equation in the presence of an electromagnetic field is given by:
\begin{eqnarray}\label{K-G-WL}
\left[ (i\partial-eA)^{2}-m_{p}^{2}\right]\psi_{p}(x)=0,
\end{eqnarray}
with $ m_{p} $ may be either the mass $ m_{K^{+}} $ of the charged kaon or $ m_{\pi^{+}} $ of the charged pion. The equation (\ref{K-G-WL}) admits as solution the following Volkov functions:
\begin{eqnarray}\label{K-G_Volckov_solution}
\psi_{K^{+}}(x)=\dfrac{1}{\sqrt{2Q_{1}V}}e^{iS(q_{1},x)},\qquad \text{and} \qquad\psi_{\pi^{+}}(x)=\dfrac{1}{\sqrt{2Q_{2}V}}e^{iS(q_{2},x)},
\end{eqnarray}
where:
\begin{eqnarray}
S(q_{j},x)=q_{j}.x-e\dfrac{a_{1}.P_{j}}{k.P_{j}}\sin\phi+e\dfrac{a_{2}.P_{j}}{k.P_{j}}\cos\phi,
\end{eqnarray}
$ q_{j}=(Q_{j},\vec{q_{j}}) $ ($ j=\{1,2\} $) is the four-momentum of the charged particle in the presence of the laser field with $Q_{j}$ denotes its effective energy. It is related to its corresponding free momentum by: $ q_{j}=P_{j}-[(e^{2}a^{2})/(2k. P_{j})]k $, such that $ {q_{j}}^{2}={P_{j}}^{2}-e^{2}a^{2}=m_{p}^{2}-e^{2}a^{2}= {m_{p}^{*}}^{2} $, where $ m_{p}^{*} $ is the effective mass of the charged particle.
In the presence of the electromagnetic field, the matrix element $ S'_{fi} $ in the first Born approximation is given by the following expression:
\begin{eqnarray}\label{smatrix_Laser_1}
S'_{fi}& = &\dfrac{G_{F}}{\sqrt{2}}\int d^{4}x\dfrac{F_{\pi^{+}}}{\sqrt{8Q_{1}Q_{2}E_{3}V^{3}}}\left[ f_{+}(\Delta')[q_{1}+P_{3}]_{\mu}q_{2}^{\mu}+f_{-}(\Delta')[q_{1}-P_{3}]_{\mu}q_{2}^{\mu}\right] \nonumber \\
& \times &e^{i(-S(q_{1},x)+ S(q_{2},x)+ P_{3}.x)},
\end{eqnarray}
where $ \Delta'$ is the new 4-momentum transfer.
To expand the element $ S'_{fi} $, we will perform some transformations in order to find an element expandable into a series of ordinary Bessel functions, with argument $z$ and phase $\phi_{0}$ such that:
\begin{eqnarray}
 z=\sqrt{\alpha_{1}^{2}+\alpha_{2}^{2}} \qquad \text{and} \qquad  \phi_{0}=\arctan(\dfrac{\alpha_{2}}{\alpha_{1}}),
\end{eqnarray}
where:
\begin{eqnarray}
\alpha_{1}=e(\dfrac{a_{1}.P_{1}}{k.P_{1}}-\dfrac{a_{1}.P_{2}}{k.P_{2}}) \qquad \text{and} \qquad \alpha_{2}=e(\dfrac{a_{2}.P_{1}}{k.P_{1}}-\dfrac{a_{2}.P_{2}}{k.P_{2}}),
\end{eqnarray}
The exponential term in equation (\ref{smatrix_Laser_1}) takes the following form:
\begin{eqnarray}\label{Action_transformations}
-S(q_{1},x)+S(q_{2},x)+P_{3}.x =-(q_{1}-q_{2}-P_{3}).x+z\sin(\phi-\phi_{0}).
\end{eqnarray}
The decay matrix element can be developed using the following ordinary Bessel transformation:
\begin{eqnarray}\label{Bessel_transformations}
 e^{iz\sin(\phi-\phi_{0})}=\sum_{n=-\infty}^{n=+\infty}J_{n}(z)e^{-in\phi_{0}}e^{in\phi}=\sum_{n=-\infty}^{n=+\infty}\boldsymbol{B}n(z)e^{in\phi}.
\end{eqnarray}
After some substitutions using the equations (\ref{Action_transformations}) and (\ref{Bessel_transformations}), the matrix element $ S'_{fi} $ becomes:
\begin{eqnarray}\label{smatrix_Laser_2}
S'_{fi}&=&\sum_{n=-\infty}^{n=+\infty}\dfrac{G_{F}F_{\pi^{+}}(2\pi)^{4}\delta^{4}(q_{2}+P_{3}-q_{1}+nk)}{4\sqrt{Q_{1}Q_{2}E_{3}V^{3}}}\left[  f_{+}(\Delta')[q_{1}+P_{3}]_{\mu}q_{2}^{\mu}+f_{-}(\Delta')[q_{1}-P_{3}]_{\mu}q_{2}^{\mu}\right] \boldsymbol{B}n(z) \nonumber \\
&=&\sum_{n=-\infty}^{n=+\infty}S'^{n}_{fi}.
\end{eqnarray}
Following the same steps as in the absence of the laser field, the decay width takes the following form:
\begin{eqnarray}\label{partial_decay}
\Gamma &=& \sum_{n=-\infty}^{n=+\infty} \dfrac{G_{F}^{2}F_{\pi^{+}}^{2}}{64\pi^{2} Q_{1}}\int \dfrac{d^{3}q_{2}}{Q_{2}}\int \dfrac{d^{3}P_{3}}{E_{3}}\delta^{4}(q_{2}+P_{3}-q_{1}+nk)\vert\mathcal{M}_{fi}^{n}\vert^{2}=\sum_{n=-\infty}^{n=+\infty}\Gamma^{n} \nonumber\\
&=& \sum_{n=-\infty}^{n=+\infty} \dfrac{G_{F}^{2}F_{\pi^{+}}^{2}}{64\pi^{2} Q_{1}}\int d\vert \vec{q}_{2}\vert F(\vert \vec{q}_{2}\vert)\delta(G(\vert \vec{q}_{2}\vert)),
\end{eqnarray}
where $ \mathcal{M}_{fi}^{n} $ represents the laser-assisted decay amplitude, and it is given by:
\begin{eqnarray}
\vert\mathcal{M}_{fi}^{n}\vert^{2}=\left[ q_{1}q_{2}\left( f_{+}(\Delta')+f_{-}(\Delta') \right) +q_{2}P_{3}\left( f_{+}(\Delta')-f_{-}(\Delta') \right) \right]^{2},
\end{eqnarray}
with $ \Delta'={m_{K^{+}}^{*}}^{2}+m_{\pi^{0}}^{2}-2q_{1}P_{3} $ is the 4-momentum transfer in the presence of the laser field. The functions $ F(\vert \vec{q}_{2}\vert) $ and $ G(\vert \vec{q}_{2}\vert) $ are defined in the appendix.
Since the charged kaon may decay via fifty decay modes\cite{PDG:2020}, and since our purpose is to study the effect of the laser field on both the hadronic channel $\{\pi^{+},\pi^{0}\}$ and the leptonic channel, we define the total decay width of the charged kaon $\Gamma^{T}$ as follows:
\begin{eqnarray}
\Gamma^{T}=\Gamma+\Gamma^{Lep.}+\Gamma^{O.C.},
\end{eqnarray}
where  $\Gamma$ is the two-body hadronic decay width given by (\ref{partial_decay}), $\Gamma^{Lep}$ represents the laser-assisted leptonic decay width of the positive kaon, and $ \Gamma^{O.C.}=8.38365\times10^{-9}\,eV $ is the sum of decay widths which correspond to other channels.
The branching ratio $ BR $ and the lifetime $ \tau $ are defined as:
\begin{eqnarray}\label{BR_et_LT}
BR=\dfrac{\Gamma}{\Gamma^{T}}, \qquad \qquad \tau=\dfrac{1}{\Gamma^{T}}.
\end{eqnarray}
Experimentally the ratio of the partial free-decay width of $ K^{+}\longrightarrow\pi^{+}+\pi^{0} $ to that of the leptonic free-decay $ K^{+}\longrightarrow\mu^{+}+\nu_{\mu} $ of positive kaon is equal to $ 0.3325\pm0.0032 $\cite{PDG:2020}.
\section{Results and discussion}\label{Sec2}
\begin{figure}[H]
\centering
  \begin{minipage}[t]{0.44\textwidth}
  \centering
    \includegraphics[width=\textwidth]{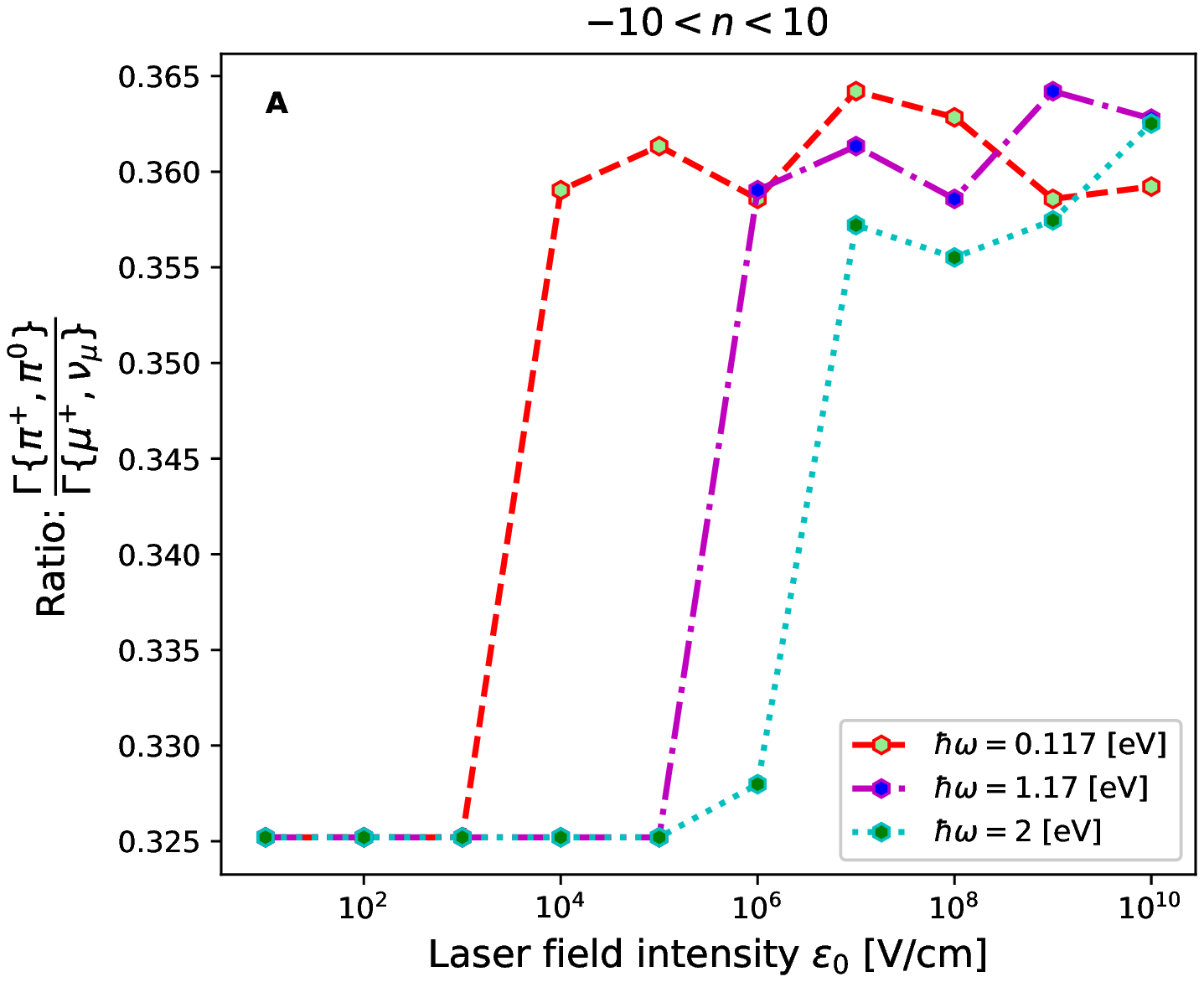}
  \end{minipage} %
  \hspace*{0.25cm}
  \begin{minipage}[t]{0.49\textwidth}
  \centering
    \includegraphics[width=\textwidth]{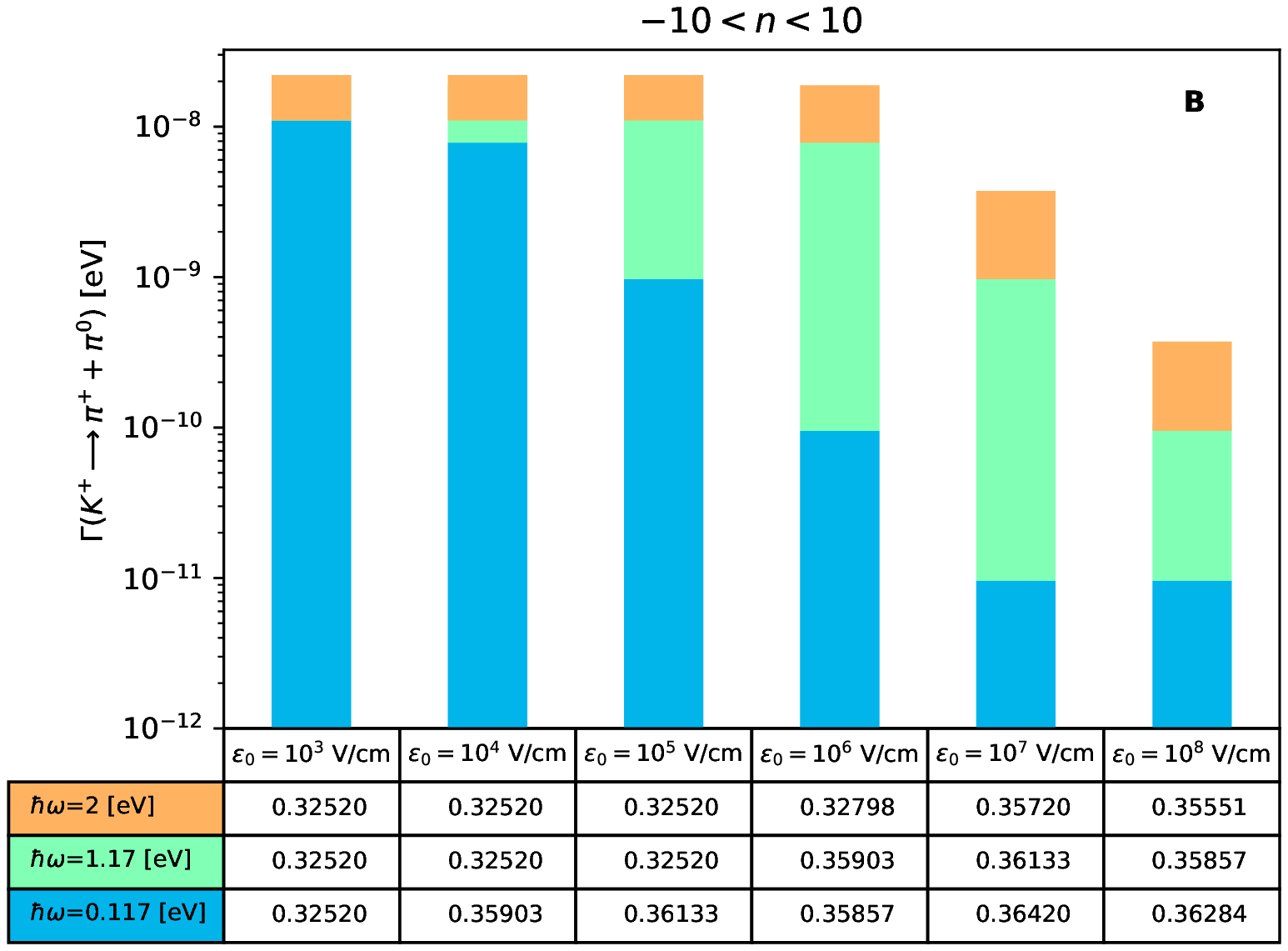}
  \end{minipage}
\caption{\textbf{A}: Ratio of the two-body charged kaon haronic decay width to the muonic decay width as a function of the laser field strength, $\varepsilon_{0}$, for different values of $\hbar\omega$ and by summing over $n$ from $-10$ to $10$. \textbf{B}: The decay width of $ K^{+} \rightarrow\pi^{+}+\pi^{0} $ as a function of the laser field strength for different values of $ \hbar\omega$ and by summing over $n$ from $-10$ to $10$.}\label{Figure:3}
\end{figure}

After the theoretical treatment of the laser-assisted two-body hadronic decay of the charged kaon, we will now discuss the different obtained results about the action of a monochromatic and discrete electromagnetic field with circular polarization on this decay. To do this, we will represent theses results in a spherical geometry chosen such that the angle $ \varphi $ associated with the produced particle $ \pi^{+}$ be zero, and the wave vector $ \vec{k} $ of the laser field propagates along the direction of the $z$-axis. We have calculated the parameters $ f_{+}(\Delta) $ and $ f_{-}(\Delta) $ in the absence of the laser field by using the value of $ f_{+}(0)=0.982\pm0.008 $\cite{PDG:2020} in the parameterization $\{ \lambda_{+},\lambda_{0}\} $. The constant $ (G_{F}\times F_{\pi^{+}})^{2} $ is obtained by normalizing the free decay width, expressed in equation (\ref{Decay_1}), by its experimental value $1.09897\times10^{-8}\,eV$\cite{PDG:2020}.
As a first step, we will present the effect of the electromagnetic field on the number of emissions and absorptions of photons $n$ in order to introduce the notion of cut-off.
\begin{figure}[H]
\centering
  \begin{minipage}[t]{0.48\textwidth}
  \centering
    \includegraphics[width=\textwidth]{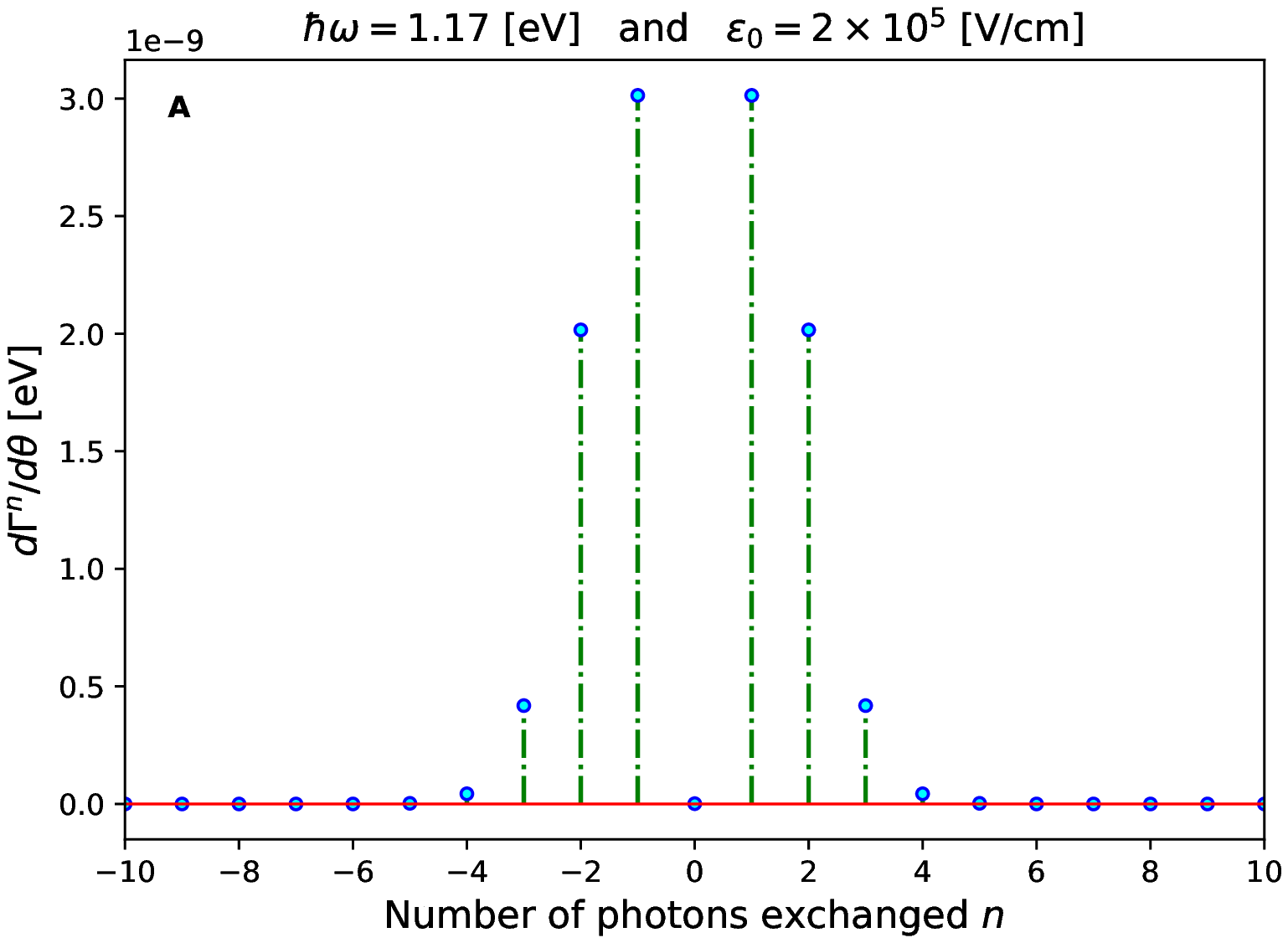}
  \end{minipage}
  \hspace*{0.25cm}
  \begin{minipage}[t]{0.48\textwidth}
  \centering
    \includegraphics[width=\textwidth]{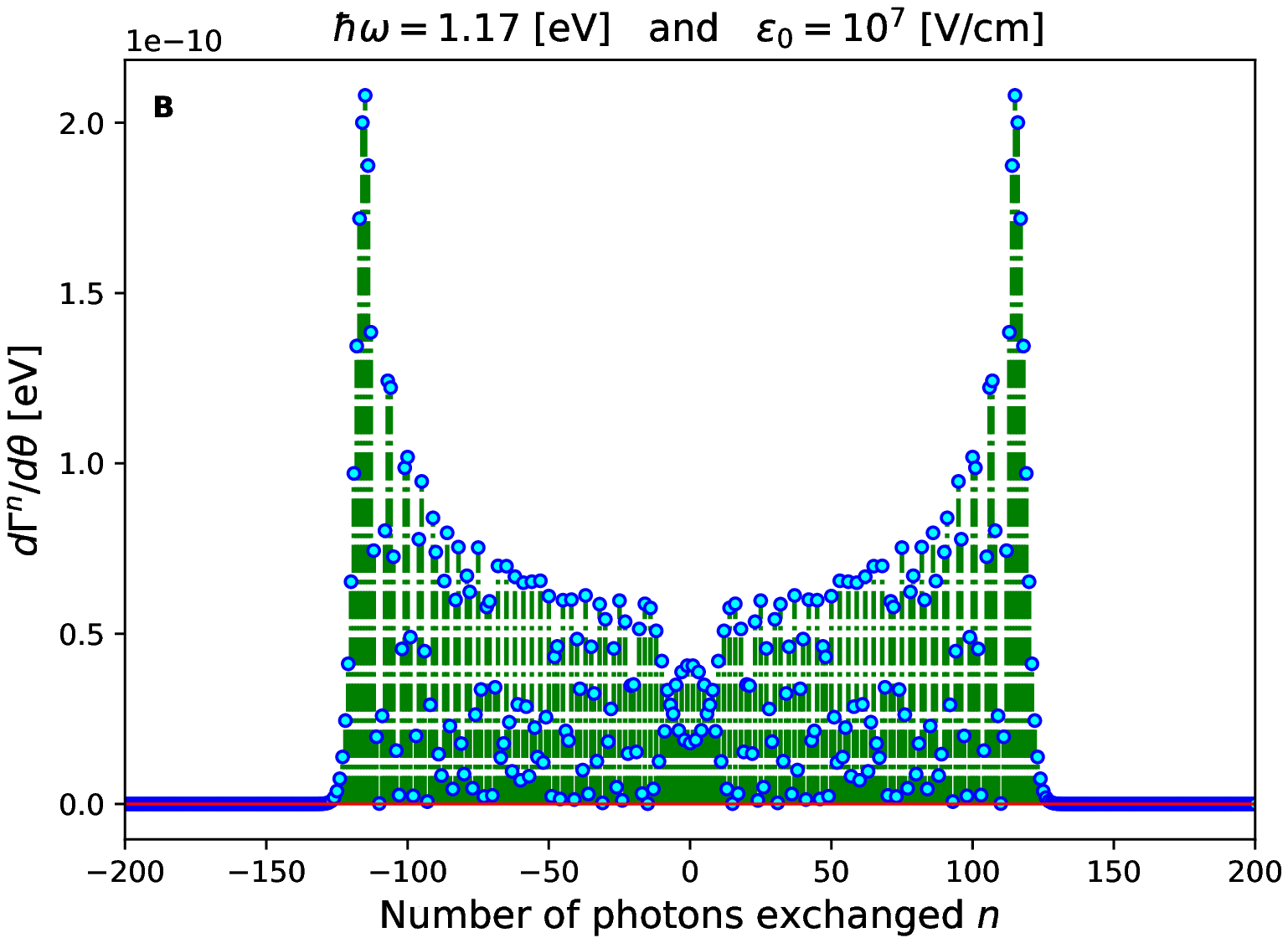}
  \end{minipage}
   \begin{minipage}[t]{0.48\textwidth}
  \centering
    \includegraphics[width=\textwidth]{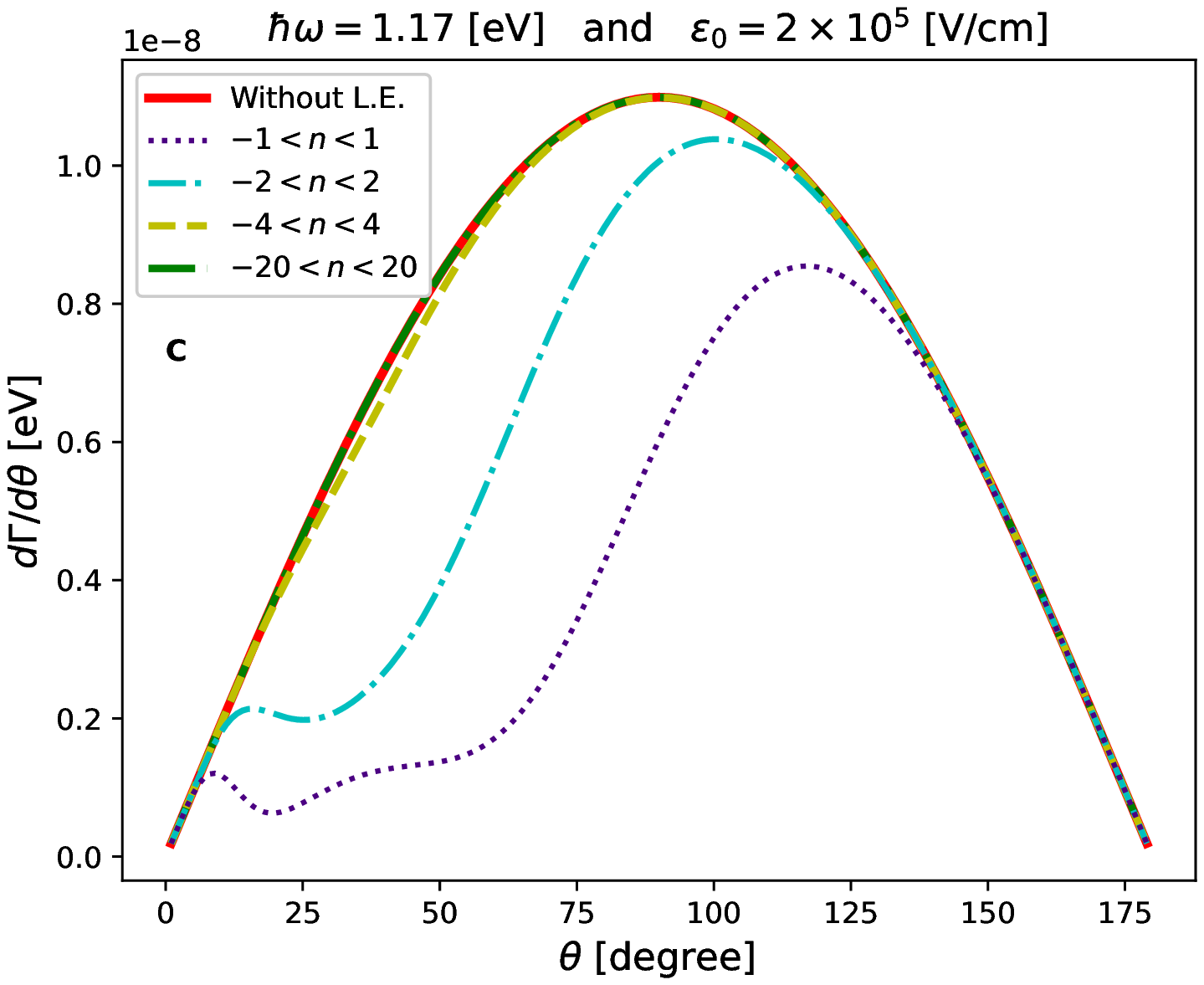}
  \end{minipage}
  \hspace*{0.25cm}
  \begin{minipage}[t]{0.48\textwidth}
  \centering
    \includegraphics[width=\textwidth]{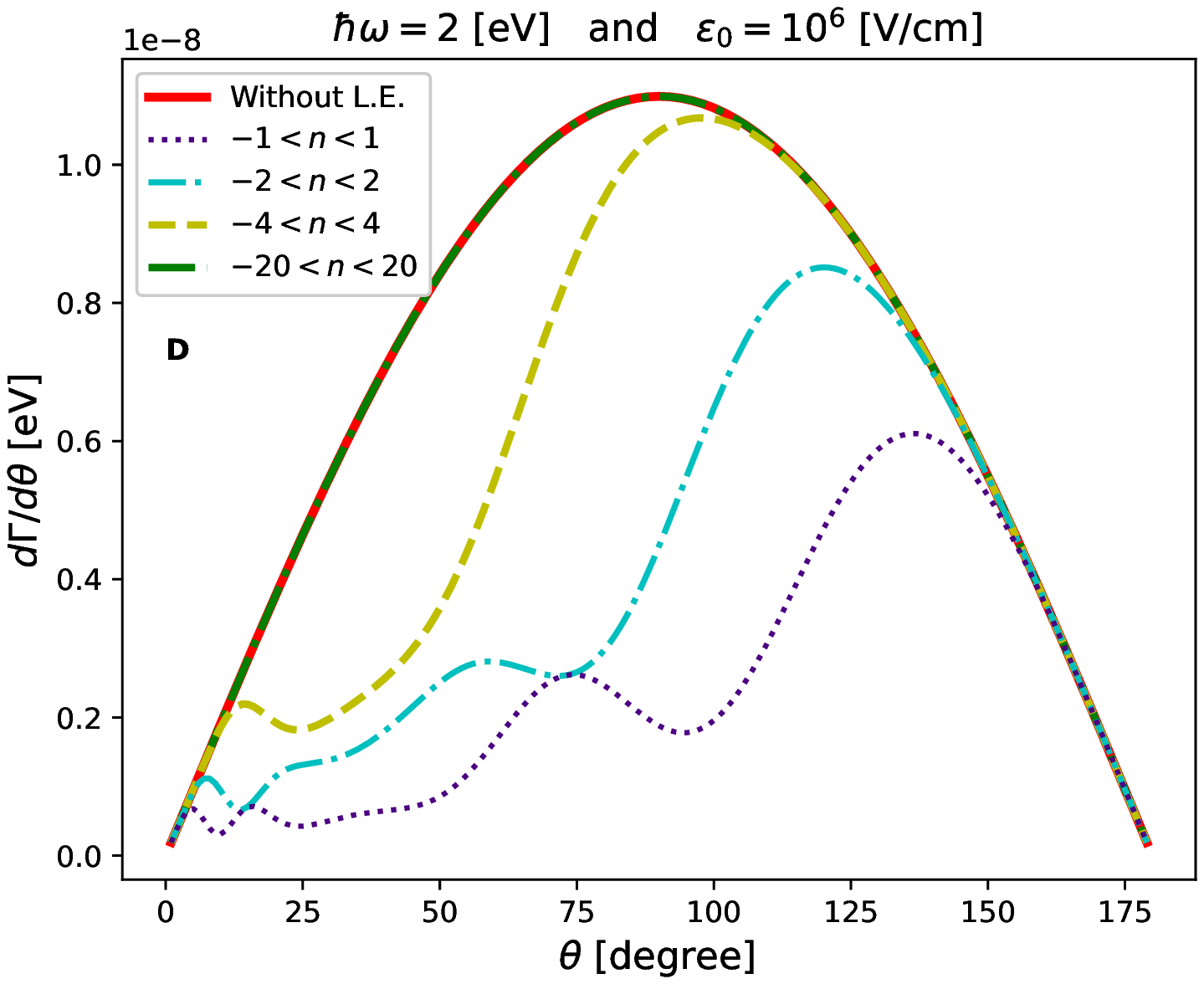}
  \end{minipage}
  \caption{\textbf{A} and \textbf{B} (top): Differential partial decay width of the charged kaon, $ \Gamma^{n}( K^{+} \rightarrow\pi^{+}+\pi^{0}) $ (\ref{partial_decay}), as a function of the photons number $n$ for $\hbar\omega=1.17\,eV$ and for different values of $\varepsilon_{0}$. The spherical coordinates are chosen such that $ \theta=90^{\circ} $ and $ \varphi=0^{\circ} $. \textbf{C} and \textbf{D} (bottom): Dependence of the laser-assisted differential decay width on the angle $\theta$ for different numbers of exchanged photons $n$ and for different values of $\hbar\omega$ and $\varepsilon_{0}$. } \label{Figure:1}
\end{figure}
According to figure \ref{Figure:1}, we observe that the variation of the laser field parameters induces a variation of the partial decay width, $d\Gamma^{n}/d\theta$, as a function of the number of photons transferred between the laser field and the decaying system. We start with the figure (\ref{Figure:1}-A), where the dressing of the charged particles has been made by a Nd:YAG laser ($\hbar\omega=1. 17\,eV$) with a strength $\varepsilon_{0}=2\times10^{5}\,V/cm$. In this case, the number of possible photons to be exchanged is $n=3$. However, when the strength of this laser (Nd:YAG laser) is increased to reach $10^{7}\,V/cm$ as it is shown in the figure (\ref{Figure:1}-B), the possible number of photons $n$ that can be exchanged may be more than $100$ photons. Outside of this range of photons numbers, we can see that the value of $d\Gamma^{n}/d\theta$ becomes zero. This means that the integral of $d\Gamma^{n}/d\theta$ over the angle $\theta$, formed by the direction of the outgoing particle $\pi^{+}$ with the $z$-axis (for a value of $\theta$ equal to $90^{\circ}$), becomes a constant for any number of photons $n$ greater than the possible number of photons that can be exchanged. The maximum number of photons that can be exchanged (absolute value of $n$) represents the notion of "cut-off" for the laser field applied with a well defined intensity. Moreover, for a number of photons which is higher than the cut-off, the obtained constant value of $\Gamma$ represents also the value of the decay width in the absence of an external field.
To illustrate more clearly these results, we have shown in figures (\ref{Figure:1}-C) and (\ref{Figure:1}-D) the effect of the number of transferred photons $n$ on the differential decay width $d\Gamma/d\theta$ for a well-defined laser field strength and by varying the angle $\theta$ from $1^{\circ}$ to $179^{\circ}$. The results presented in these figures are in full agreement with that presented in figures (\ref{Figure:1}-A) and (\ref{Figure:1}-B). Indeed, the differential decay width in the presence of the laser field coincides with that in the absence of the laser field (red curve in figure (\ref{Figure:1}-C) for $\theta=90^{\circ}$) when we sum over $n$ from $-4$ to $4$ or from $-20$ to $20$. In addition, we observe that in the case of a summation over $n$ from $-20$ to $20$ in both figure (\ref{Figure:1}-C) and (\ref{Figure:1}-D), the laser-assisted differential decay width (green curve) is equal to its corresponding laser-free differential decay width for all values of $\theta$ between $1^{\circ}$ and $179^{\circ}$. This result indicates that $\vert n\vert=20$ exceeds the cut-off which corresponds to both ($\varepsilon_{0}=2\times10^{5}\,V/cm$; $\hbar\omega=1.17\,eV$) and ($\varepsilon_{0}=10^{6}\,V/cm$; $\hbar\omega=2\,eV$).
In the following section, we will focus on the impact of the laser field on both the branching ratio and the lifetime, which are expressed in the equation (\ref{BR_et_LT}).
 \vspace{-0.5cm}
\begin{figure}[H]
\centering
  \begin{minipage}[t]{0.48\textwidth}
  \centering
    \includegraphics[width=\textwidth]{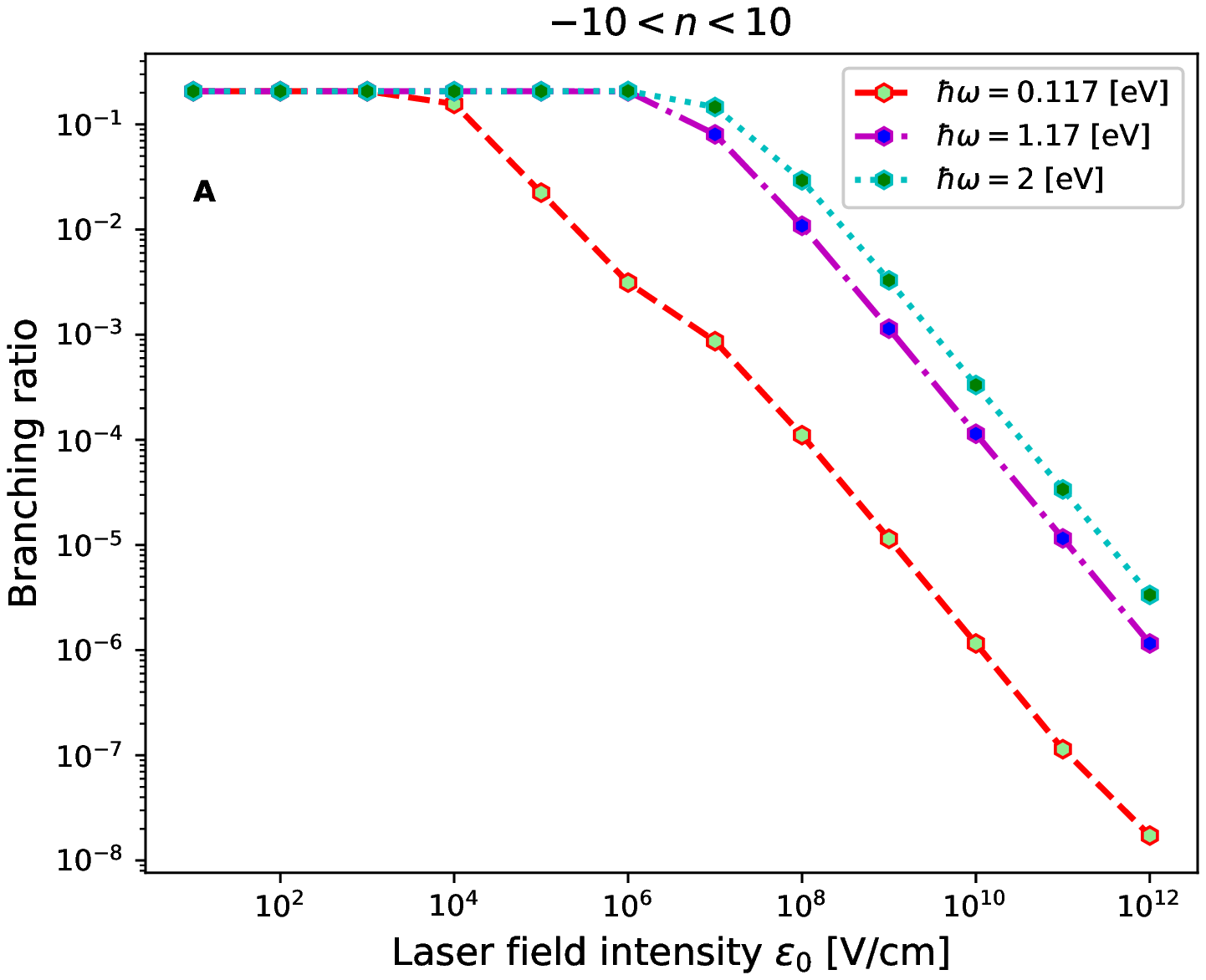}
  \end{minipage}
  \hspace*{0.25cm}
  \begin{minipage}[t]{0.48\textwidth}
  \centering
    \includegraphics[width=\textwidth]{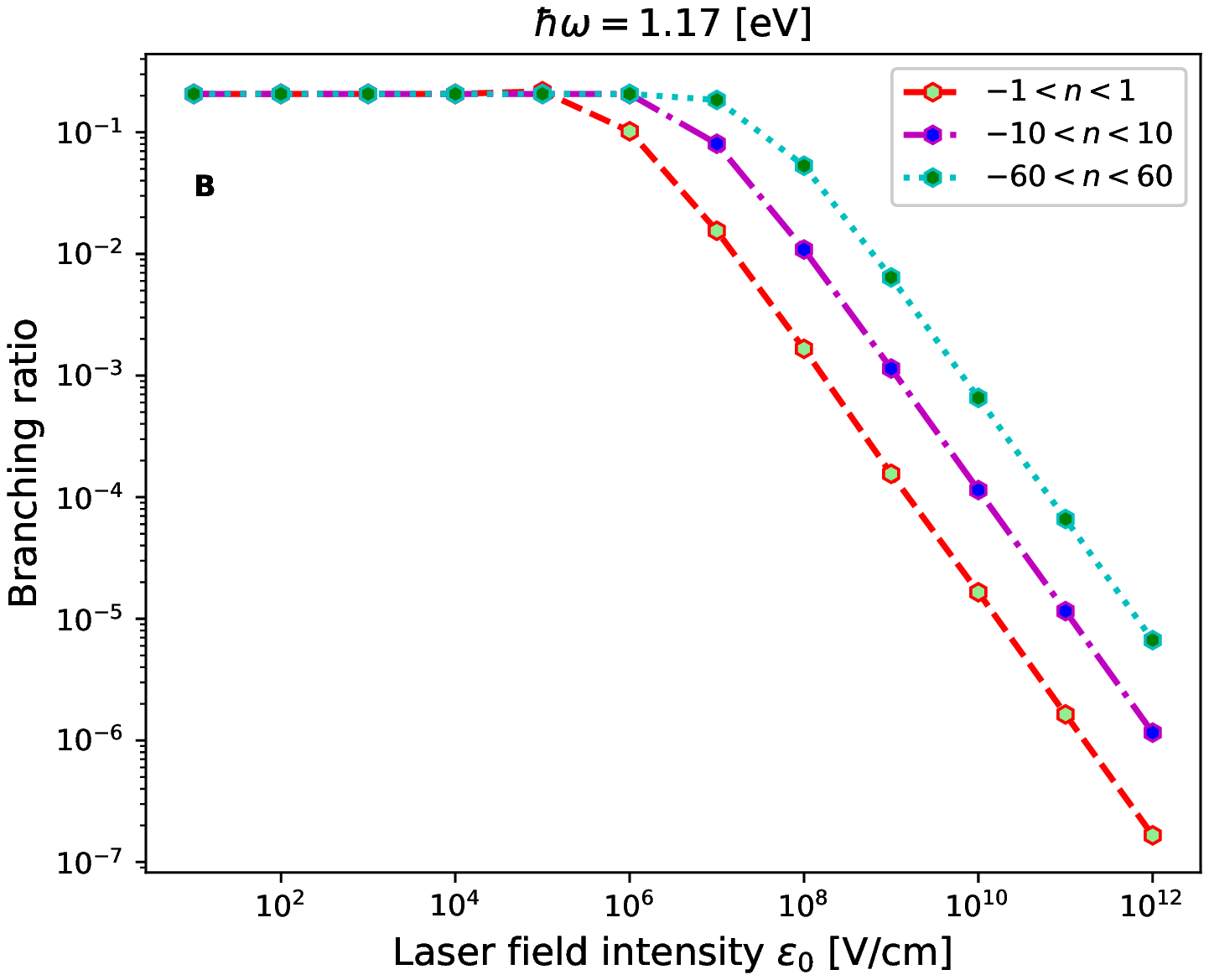}
  \end{minipage}
   \begin{minipage}[t]{0.48\textwidth}
  \centering
    \includegraphics[width=\textwidth]{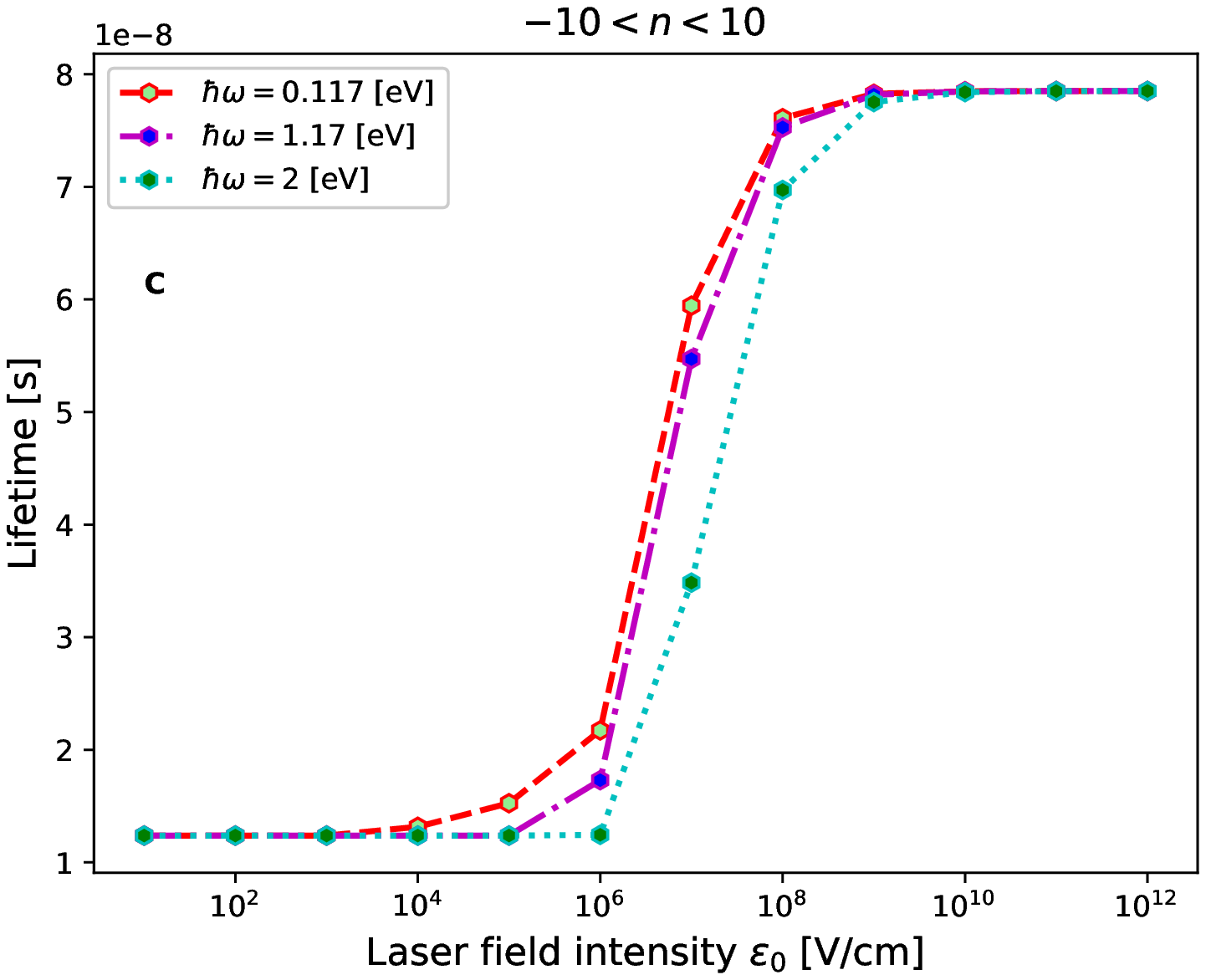}
  \end{minipage}
  \hspace*{0.25cm}
  \begin{minipage}[t]{0.48\textwidth}
  \centering
    \includegraphics[width=\textwidth]{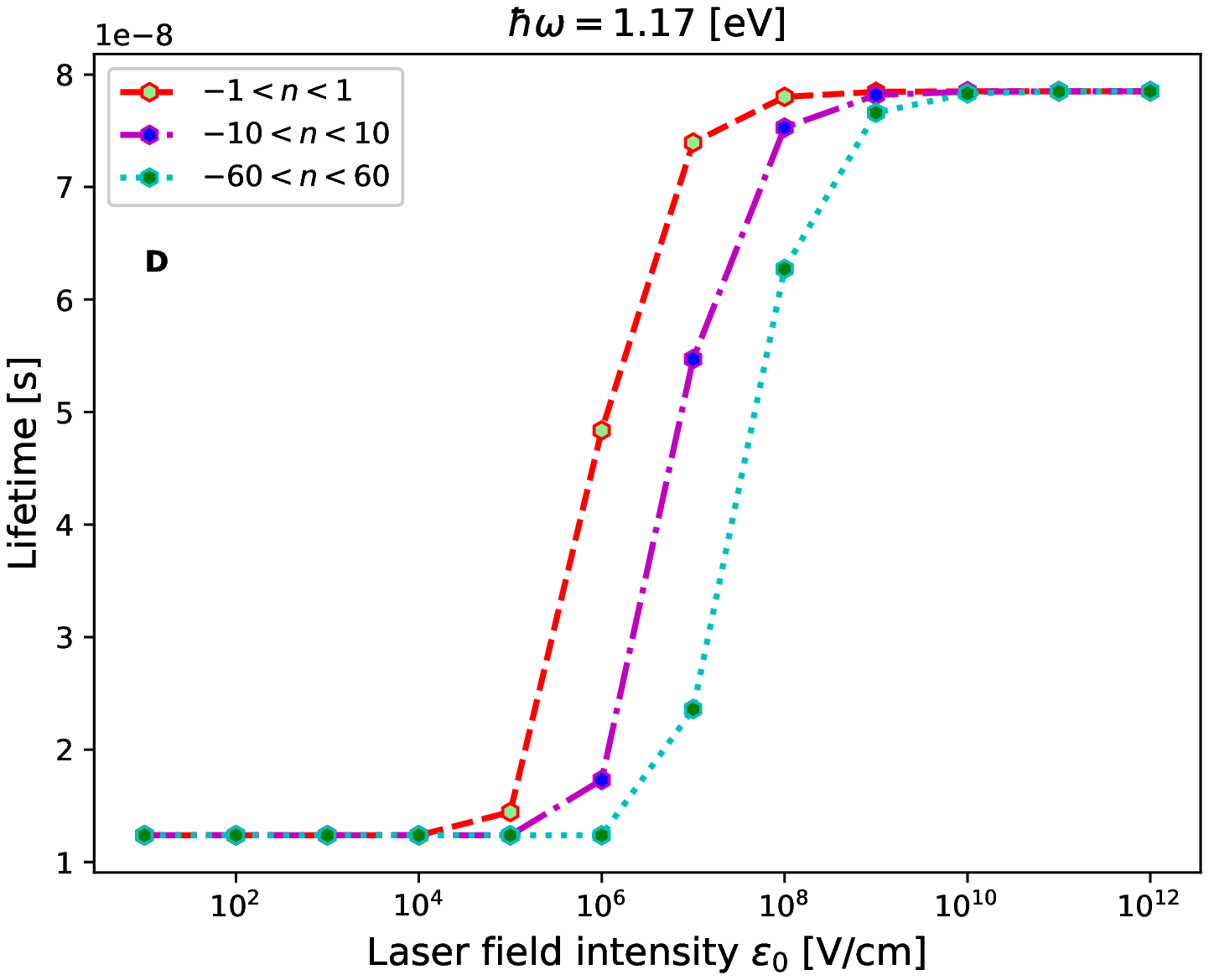}
  \end{minipage}
  \caption{\textbf{A} and \textbf{B} (top): Branching ratio of the charged kaon decay as a function of the laser field strength $\varepsilon_{0}$. \textbf{A}: for different values of $ \hbar\omega $ and by summing over $n$ from $-10$ to $10$. \textbf{B}: for different values of $n$ and for $ \hbar\omega=1.17\,eV$. \textbf{C} and \textbf{D} (bottom): Lifetime of the positive kaon as a fucntion of the laser field strength. \textbf{C}: for different known laser frequencies and by taking the number of exchanged photons as ranging from $-10\leq n\leq 10$. \textbf{D}: for different values of $n$ and for $ \hbar\omega=1.17\,eV$.} \label{Figure:2}
\end{figure}
Figure (\ref{Figure:2}-A) represents the variation of the branching ratio as a function of the laser field strength for different frequencies $\hbar\omega=\{0.117\,eV,\, 1.17\,eV\, \text{and}\, 2\,eV\}$ and by summing over the exchanged photons number from $-10$ to $10$. It is well known that in the absence of the laser field the branching ratio is experimentally equal to $(20.67\pm0.08)\times10^{-2}$\cite{PDG:2020}. According to this figure (\ref{Figure:2}-A), we observe that, for low intensities, the laser field doesn't affect the branching ratio. However, the branching ratio begins to decrease progressively from the laser field strength $\varepsilon_{0}=10^{3}\,V/cm$ for $\hbar\omega=0.117\,eV$.
For the $CO_{2}$ laser ($\hbar\omega=0.117\,eV$), and for $\varepsilon_{0}=10^{12}\,V/cm$, the branching ratio is equal to $1. 686\times10^{-8}\,s$. Moreover, with this strength value, we mention that the decay process can exchange more than $100000$ photons. Therefore, the interpretation of the branching ratio decrease is more significant for strengths in the order of $10^{4}\,V/cm$.  For the two other lasers, the branching ratio begins to decrease from $\varepsilon_{0}=10^{5}\,V/cm$ for a Nd:YAG laser ($\hbar\omega=1.17\,eV$), and also from the $\varepsilon_{0}=10^{6}\,V/cm$ for the He:Ne laser ($\hbar\omega=2\,eV$).
In the figure (\ref{Figure:2}-B) and for the Nd:YAG laser, the branching ratio always decreases by increasing the laser field strength, regardless of the number of exchanged photons $n$.
The cases where the branching ratio is constant corresponds to the strengths (low laser field strengths) for which the process can only exchange a number of photons $\vert n\vert$ lower than that shown in the figure. This means that the number of photons exchanged exceeds the cut-off.
Now, let's move to the figures (\ref{Figure:2}-C) and (\ref{Figure:2}-D) that illustrate the variation of the laser-assisted lifetime of the charged kaon. It is clear that the lifetime increases as a function of the laser field strength especially when its strength overcome a threshold value.
Then, it stagnates as the laser field strength reaches a certain value.
As we have seen for the branching ratio, the lifetime of the charged kaon (figure (\ref{Figure:2}-C)) in the presence of the laser field is more significant for the laser field strengths $\varepsilon_{0}=10^{4}\,V/cm$, $\varepsilon_{0}=10^{6}\,V/cm$ and $\varepsilon_{0}=10^{7}\,V/cm$ and for the frequencies $\hbar\omega=0. 117\,eV$, $\hbar\omega=1.17\,eV$ and $\hbar\omega=2\,eV$, respectively. By comparing these results with those obtained in the case of laser-assisted leptonic decay, we notice that the branching ratio and lifetime of the laser-assisted two-body hadronic decay of the charged kaon behaves in the same manner as in the case of leptonic dacy ($K^{+}\longrightarrow\mu^{+}+\nu_{\mu}$). However, we remark that the lifetime is longer in the case of  the hadronic dressing channel\cite{Baouahi:2021}. In figure (\ref{Figure:3}) the ratio of the partial decay width of $K^{+}\longrightarrow\pi^{+}+\pi^{0}$ to that of $K^{+}\longrightarrow\mu^{+}+\nu_{\mu}$ is considered.

Figure (\ref{Figure:3}-A) illustrates the ratio of the $K^{+} \rightarrow\pi^{+}+\pi^{0}$ to that of $K^{+} \rightarrow\mu^{+}+\nu_{\mu}$ for different known laser frequencies and by taking $n$ as ranging from  $-10$ to $10$. We recall that the experimental ratio of the two decay widths is equal to $0.3252\pm0.0016$\cite{PDG:2020}, and the branching ratio in the muonic channel is equal to $(63.56\pm0.11)\%$ in the absence of the laser. We also mention that the  results of the two-body muonic decay of the charged kaon in the presence of a laser field are taken from our previous work\cite{Baouahi:2021}.
According to figure (\ref{Figure:3}-A), it is clear that, for different laser field frequencies, the studied ratio undergoes an augmentation especially for the laser field strengths that allow the exchange of a number $\vert  n\vert=10$ of photons (e.g. the strengths which are greater than $\varepsilon_{0}=\{10^{3}\,V/cm$ for the case of $\hbar \omega=0.117\,eV$ ). As an hypothesis, the increase of the ratio of the hadronic decay width to the muonic decay width may be interpreted by the fact that the laser field increases the two-body hadronic decay width of the charged kaon, and decreases its muonic decay width. Another hypothesis is that both the hadronic and muonic decay width increase or decrease, but they change with different rhythms. To check the validity of our hypotheses, we have presented different values of the ratio $\Gamma\{\pi^{+},\pi^{0}\}/\Gamma\{\mu^{+},\nu_{\mu}\}$ in the table associated to the figure (\ref{Figure:3}-B) to illustrate the variation of the decay width of $K^{+} \rightarrow\pi^{+}+\pi^{0} $ as a function of the laser field strength.
As presented in figure (\ref{Figure:3}-B), the two-body hadronic decay width of the charged kaon decreases by increasing the laser field strength. This means that our second hypothesis is the correct one. Therefore, the laser field has a strong effect of decreasing the charged kaon muonic decay width, $K^{+}\longrightarrow\mu^{+}+\nu_{\mu}$, as compared to its effect on the hadronic decay width presented in figure (\ref{Figure:3}-B). We should mention that in the histogram which corresponds to $\hbar \omega=2\,eV$ and for the intensities $\varepsilon_{0}=\{10^{3}\,V/cm,\,10^{4}\,V/cm,\, 10^{5}\,V/cm\,     \text{and}\, 10^{6}\,V/cm\}$ the decay width has different values others than the ones represented in (\ref{Figure:3}-B). The real value of this decay width are successively $\Gamma=1.09897\times10^{-8}\,eV$ and $\Gamma=1. 09894\times10^{-8}\,eV$ for the intensities $\varepsilon_{0}=\{10^{3}\,V/cm,\,10^{4}\,V/cm,\, 10^{5}\,V/cm\}$ and $\varepsilon_{0}=10^{6}\,V/cm$. This change is made to illustrate clearly the decrease of the decay width. After studying the effect of the circularly polarized laser field on various measurable quantities, we will now move to the last obtained result which represents the simultaneous dependence of the charged kaon decay width on both the laser field strength and the number of exchanged photons for $\theta=90^{\circ}$.
\begin{figure}[H]
\centering
  \begin{minipage}[t]{0.48\textwidth}
  \centering
    \includegraphics[width=\textwidth]{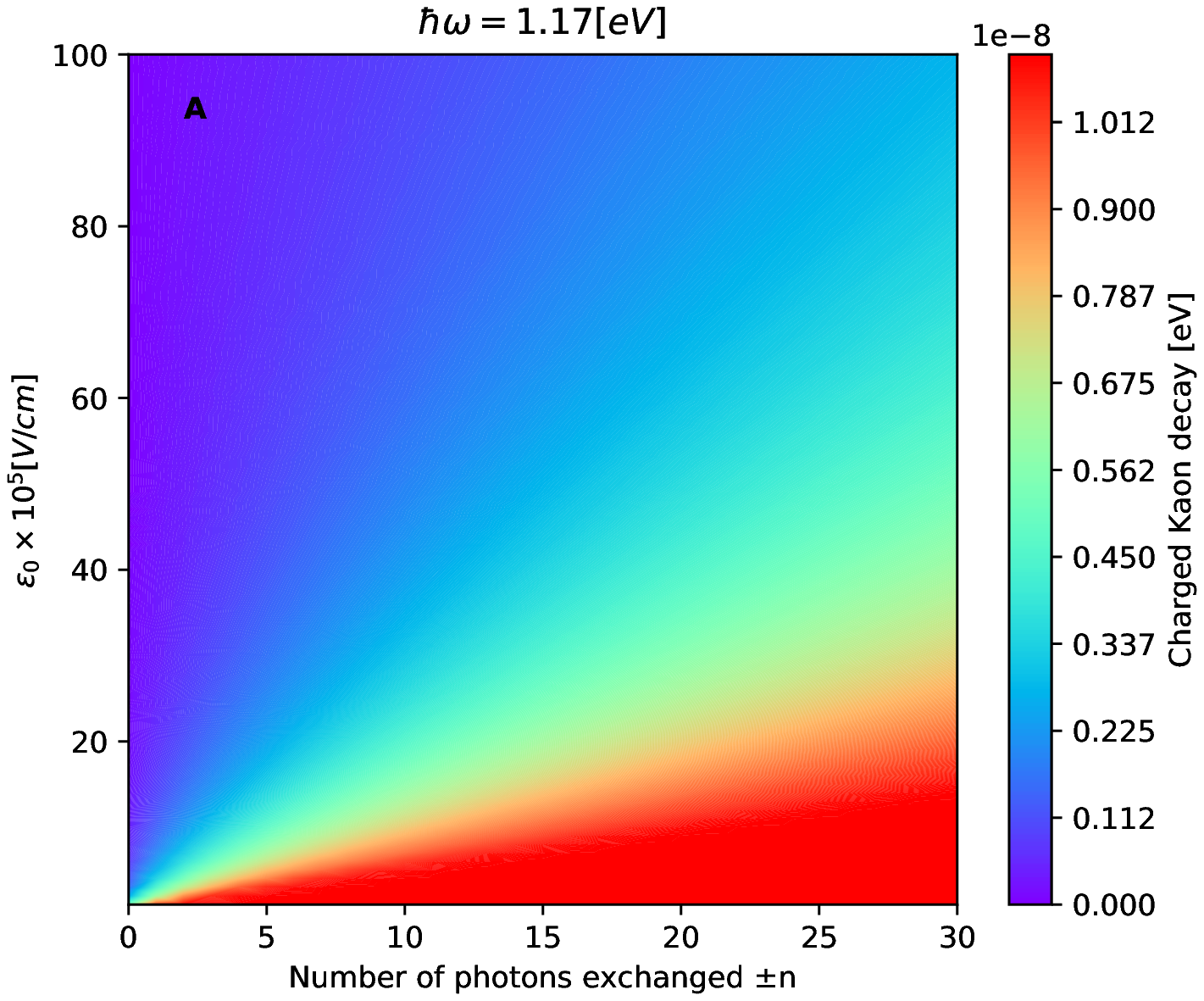}
  \end{minipage} %
  \hspace*{0.25cm}
  \begin{minipage}[t]{0.48\textwidth}
  \centering
    \includegraphics[width=\textwidth]{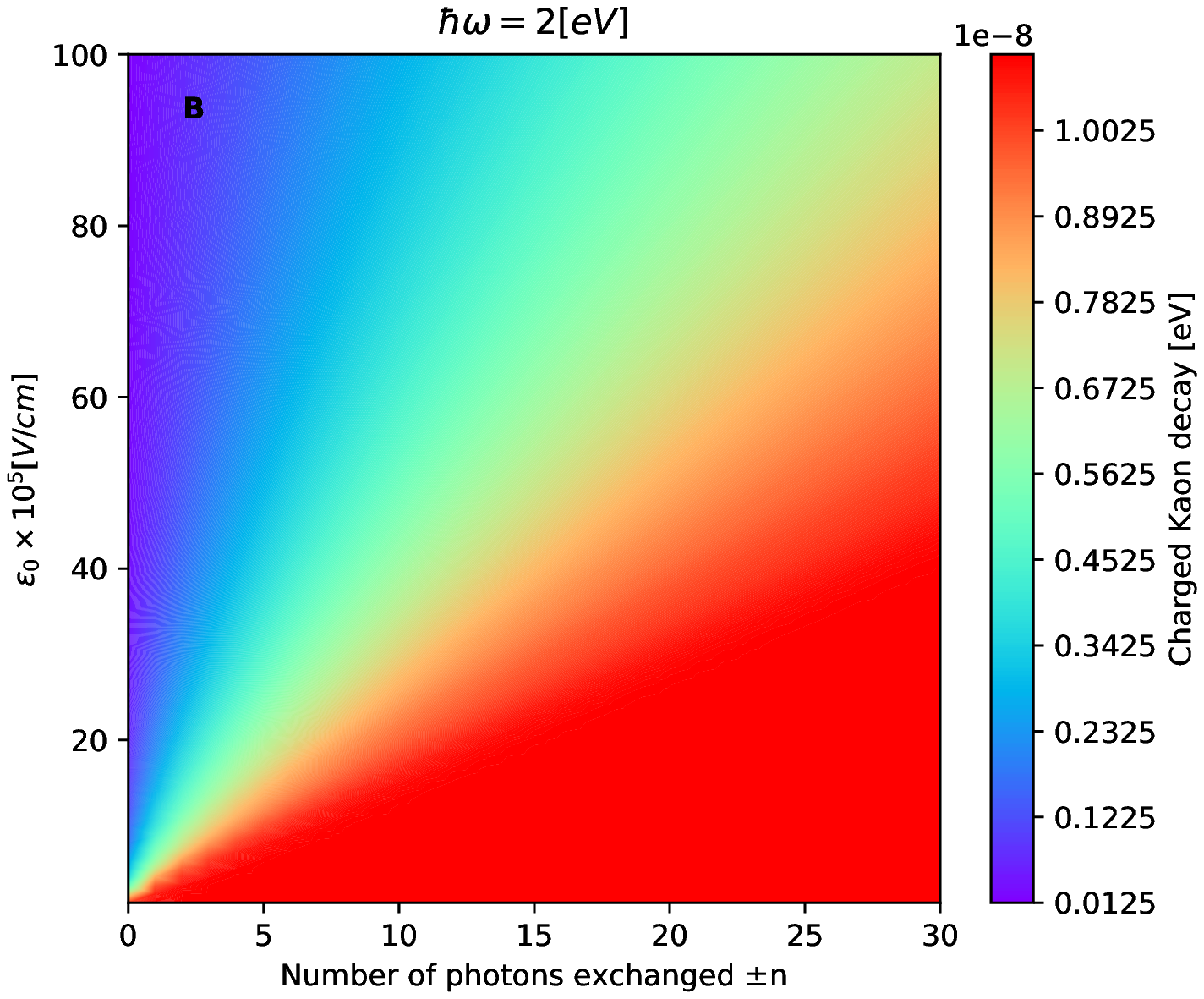}
  \end{minipage}
\caption{Variation of the charged kaon decay width for different values of the laser field strength and for different ranges of transferred photons $n$. Left pannel (\textbf{A}): $ \hbar\omega=1.17\,eV$, right pannel (\textbf{B}): $ \hbar\omega=2\,eV$.}\label{Figure:4}
\end{figure}
Figure \ref{Figure:4} represents some values of the charged kaon decay width for different combinations of the laser field and the number of exchanged photons. The range of laser field strengths $\varepsilon_{0}=10^{5}\,V/cm  \rightarrow \varepsilon_{0}=10^{7}\,V/cm$ have been selected  based on the results obtained in figures (\ref{Figure:2}) and (\ref{Figure:3}).
As figure \ref{Figure:4} shows, the red zone corresponds to the maximum decay width.
By comparing the results obtained in figures (\ref{Figure:1}-A) and (\ref{Figure:1}-B) with those of figures (\ref{Figure:4}-A) and (\ref{Figure:4}-B), we can deduce that the number of photons in this zone are higher than its corresponding cutoffs.
The violet zone in both figures (\ref{Figure:4}-A) and (\ref{Figure:4}-B) corresponds to values of the charged kaon decay width which are near to zero. This zone is less probable since the required photons number representing the cutoff in this zone is very large than that shown in these figures.
Therefore, we predict that the charged kaon decay width is more significant in the region between the red and violet zones. Another important point to be mentioned, here, is that by comparing the results of figure (\ref{Figure:4}-A) with those of figure (\ref{Figure:4}-B), for the same range of laser strengths, it appears that the use of high frequencies requires high laser field strengths in order to introduce the phenomena of emission and/or absorption of photons. In contrast, for the Nd:YAG laser, low laser field strengths can induce the exchange of photons with the decaying system.
\section{Conclusion}\label{Sec3}
In this paper and based on the interesting results found for laser-assisted leptonic decay \cite{Baouahi:2021}, we have investigated the charged kaon decay in the hadronic channel in the presence of a circularly polarized laser field.
By theoretically analyzing the effect of the laser field on different quantities related to the decay, we have found that the two-body positive kaon decay in the hadronic channel is modified by a classical monochromatic laser field with circular polarization. In addition, the decay width decreases by increasing the laser field strength, and the emission and/or absorption phenomena appear and increase by increasing the laser field strength. Moreover, the study of mesons as particles without structure introduces the decrease of the branching ratio with an increase in the lifetime. We have also found that the ratio of the hadronic decay width to the muonic decay width increases by increasing the strength of the laser field. This result is due to the strong and rapid decrease of the muonic decay width as compared to the two-body hadronic decay width. All these results invite us to study other channels of the charged kaon decay in the presence of the electromagnetic field in order to have a complete picture about its effect on the decay.
\appendix
\section*{Appendix}
To find the final expression of $ \Gamma $ in the equation (\ref{partial_decay}), we use the following integral property:
\begin{eqnarray}
\int F(x)\delta(G(x))dx={\dfrac{F(x_{0})}{|G'(x_{0})|}} \qquad \text{with} \qquad G(x_{0})=0,
\end{eqnarray}
where the expression of $ F $ and $ G $ are determined as:
\begin{eqnarray}
F(\vert \vec{q}_{2}\vert)&=&\dfrac{\vert\mathcal{M}_{fi}^{n}\vert^{2}\vert\vec{q}_{2}\vert^{2}}{\sqrt{\vert\vec{q}_{2}\vert^{2}+m_{\pi^{+}}^{2}}\sqrt{m_{\pi^{0}}^{2}+\vert\vec{q}_{2}\vert^{2}+\lbrace(n+\dfrac{e^{2}a^{2}}{2kP_{1}})\omega\rbrace^{2}-2(n+\dfrac{e^{2}a^{2}}{2kP_{1}})\omega\vert\vec{q}_{2}\vert\cos(\theta)}}, \\
G(\vert \vec{q}_{2}\vert)&=& n\omega-Q_{1}+\sqrt{\vert\vec{q}_{2}\vert^{2}+m_{\pi^{+}}^{2}}+\sqrt{m_{\pi^{0}}^{2}+\vert\vec{q}_{2}\vert^{2}+\lbrace(n+\dfrac{e^{2}a^{2}}{2kP_{1}})\omega\rbrace^{2}-2(n+\dfrac{e^{2}a^{2}}{2kP_{1}})\omega\vert\vec{q}_{2}\vert\cos(\theta)}. \nonumber\\
\end{eqnarray}
In this case, the expression of $ \vert\vec{q}_{2}\vert $ will be as follows:
\begin{eqnarray}
\vert\vec{q}_{2}\vert &=&\{2n\omega(m_{\pi^{0}}^{2}-{m_{\pi^{+}}^{*}}^{2}-E_{1}^{2})(kP_{1})^{2}\cos(\theta)+e^{4}a^{4}\omega^{2}E_{1}\cos(\theta)\nonumber \\
&+& 4n^{2}\omega^{2}E_{1}(kP_{1})^{2}\cos(\theta)+e^{2}a^{2}\omega(kP_{1})\cos(\theta)\times(m_{\pi^{0}}^{2}-{m_{\pi^{+}}^{*}}^{2}-E_{1}^{2}+4n\omega E_{1})\nonumber \\
&+&[(e^{2}a^{2}\omega-2(kP_{1})(E_{1}-n\omega))^{2}\times(e^{4}a^{4}\omega^{2}(E_{1}^{2}-{m_{\pi^{+}}^{*}}^{2}\sin(\theta)^{2})+2e^{2}a^{2}\omega(kP_{1})(E_{1}(m_{\pi^{0}}^{2}\nonumber \\
&-&E_{1}^{2}+2n\omega E_{1})+{m_{\pi^{+}}^{*}}^{2}(E_{1}-2n\omega\sin(\theta)^{2}))+(kP_{1})^{2}(m_{\pi^{+}}^{4}+(m_{\pi^{0}}^{2}-E_{1}^{2}+2n\omega E_{1})^{2} \nonumber \\
&-&2m_{\pi^{+}}^{2}(m_{\pi^{0}}^{2}+E_{1}^{2}-2n\omega E_{1} +2n^{2}\omega^{2}\sin(\theta)^{2})))]^{1/2}\}/\{e^{4}a^{4}\omega^{2}\sin(\theta)^{2}+4e^{2}a^{2}\omega(kP_{1})(n\omega\sin(\theta)^{2}\nonumber \\
&-&E_{1})+4(kP_{1})^{2}(E_{1}^{2}-2n\omega E_{1}+n^{2}\omega^{2}\sin(\theta)^{2})
\}.
\end{eqnarray}

\end{document}